\documentclass[twocolumn,dvipsnames]{aastex62}

\usepackage{CJK}
\usepackage{framed}

\def\dim#1{{\rm\,#1}}
\def\Msun{{\rm\,M}_\odot}

\graphicspath{{./}{figures/}}

\submitjournal{ApJ}

\shorttitle{}
\shortauthors{Zhu and Gnedin}

\begin{document}
\begin{CJK*}{UTF8}{gkai}

\title{Cosmic Reionization On Computers: Baryonic Effects on Halo Concentrations During the Epoch of Reionization}

\correspondingauthor{Hanjue Zhu (朱涵珏)}
\email{hanjuezhu@uchicago.edu}
\author[0000-0003-0861-0922]{Hanjue Zhu (朱涵珏)}
\affiliation{Department of Astronomy \& Astrophysics; 
The University of Chicago; 
Chicago, IL 60637, USA}

\author{Nickolay Y.\ Gnedin}
\affiliation{Particle Astrophysics Center; 
Fermi National Accelerator Laboratory;
Batavia, IL 60510, USA}
\affiliation{Kavli Institute for Cosmological Physics;
The University of Chicago;
Chicago, IL 60637, USA}
\affiliation{Department of Astronomy \& Astrophysics; 
The University of Chicago; 
Chicago, IL 60637, USA}

\begin{abstract}
Baryons both increase halo concentration through adiabatic contraction and expel mass through feedback processes. However, it is not well understood how the radiation fields prevalent during the epoch of reionization affect the evolution of concentration in dark matter halos. We investigate how baryonic physics during the epoch of reionization modify the structure of dark matter halos in the Cosmic Reionization On Computers (CROC) simulations. We use two different measures of halo concentration to quantify the effects. We compare concentrations of halos matched between full physics simulations and dark-matter-only simulations with identical initial conditions between $5 \leq z \leq 9$. Baryons in full physics simulations do pull matter towards the center, increasing the maximum circular velocity compared to dark-matter-only simulations. However, their overall effects are much less than if all the baryons were simply centrally concentrated indicating that heating processes efficiently counteract cooling effects. Finally, we show that the baryonic effects on halo concentrations at $z\approx5$ are relatively insensitive to environmental variations of reionization history. These results are pertinent to models of galaxy-halo connection during the epoch of reionization.
\\
\end{abstract}


\section{Introduction}\label{sec:intro}

Under the $\Lambda$CDM paradigm, primordial density fluctuations lead to the growth of dark matter halos. Smaller dark matter halos form first, and they start to accrete matter from their surroundings and merge with other dark matter halos to form successively larger structures. Galaxies form in the gravitational potential well of the dark matter halos as the infalling gas cools and turns into stars. In this scenario, baryons and dark matter are coupled and baryons may have a significant impact on the structure of the dark matter halos. For example, baryons concentrate in the center of dark matter halos because of cooling, which makes the dark matter halo more centrally concentrated as baryons pull the dark matter inward \citep{blumenthal86,gnedin04,sellwood05,gustafsson06,pedrosa09,abadi10,sommer-larsen10,gnedin11}; supernova and AGN feedback, on the other hand, could lower the central dark matter density \citep{peirani08,governato10,pontzen2012,martizzi12,brooks14}. These baryonic processes are difficult to analytically model due to their nonlinear, stochastic, and highly coupled behaviors. Therefore, radiative hydrodynamic simulations help us understand the net effect of baryonic physics on dark matter halo structures. Predictions from these numerical simulations and a detailed theoretical modeling of the halo structure are critical for correct interpretation of a wide range of astronomical observations, from stellar and gas kinematics to gravitational lensing measurements.

The most widely used model for the dark matter halo density profile has been proposed by Navarro, Frenk, and White \citep{nfw97}. Using cosmological N-body simulations, they have shown that the density profiles of dark matter halos in suites of simulations of different cosmologies can be fitted using a universal fitting function, which astrophysicists commonly refer to as the NFW profile. The NFW profile takes the form:
\begin{equation}
\rho(r)= \frac{\rho_{\mathrm{s}}}{\left(r / r_{\mathrm{s}}\right)\left(1+r / r_{\mathrm{s}}\right)^{2}},
\end{equation}
where $r_{\mathrm{s}}$ is a characteristic radius at which the density profile slope agrees with that of the isothermal profile ($\rho(r) \propto r^{-2}$), and $\rho_{\mathrm{s}}$ is the density at the scale radius, $\rho(r_{\mathrm{s}})$. Alternatively, we can rewrite $\rho_{\mathrm{s}}$ and $r_{\mathrm{s}}$ in terms of the halo virial mass and a ``concentration" parameter. This halo concentration parameter $c$ is defined as
\begin{equation}
c \equiv \frac{R_{\Delta}}{r_{\mathrm{s}}},
\end{equation} 
where $\Delta$ is the halo overdensity with respect to the critical density for a given mass definition (e.g. virial, 200c, 500c).  The larger the value of the concentration parameter, the more concentrated dark matter is within a halo. This concentration parameter is widely used to characterize dark matter halo density profiles, because knowing both the halo concentration and mass uniquely specifies the NFW halo density profile.

\citet{nfw97} has also shown that at low redshift halo concentration decreases with increasing halo mass. They interpret this trend to be a result of hierarchical structure formation. They relate halo concentration to the background density of the Universe at the halo's formation time - halos that form earlier (when the universe is denser) are more concentrated. Since more massive halos on average form later, they have on average smaller concentration. The large scatter in the halo mass assembly history leads to a large scatter in the concentration-mass ($c-M$) relation. Subsequently, many semi-analytical models have been proposed to predict and explain the concentration-mass relation. They show that halo concentrations are related to halo mass assembly histories and the concentration parameter is a function of halo mass, redshift, and cosmological parameters \citep{bullock01,wechsler02,zhao03,zhao09,ludlow14,DK15,wang20}. 

At low redshifts, the $c-M$ relation is well-studied using N-body simulations, semi-analytical models, and large-scale full volume galaxy formation simulations \citep{schaller15,chua17,beltz-mohrmann21}. However, at high redshifts, especially at $z>6$, most of the current predictions are made with N-body simulations and semi-analytical models \citep{klypin11,prada2012,dutton2014,DK15,Angel2016,hellwing2016}. Their results show a rather large discrepancy, which is not surprising because the effect of baryonic physics depends on how the physical prescriptions for galaxy formation are implemented in the simulations/models. In this regard, both the high-z $c-M$ relation and the impact of baryonic physics on dark matter halo structure are exciting areas for future studies.

Predictions of high-z ($5\leq z \leq9$) concentration-mass relation and its dependence on redshift made by the Cosmic Reionization On Computers (CROC) simulations fit in both areas of study. CROC, specifically, is a simulation of reionization that incorporates the relevant physics of cosmic reionization in sufficiently large ($>$ 100 Mpc) boxes while resolving characteristic radii of dark matter halos, so it is in a unique position to allow us to study the impact of reionization physics on the high-z concentration-mass relation. 

This paper is organized as follows. Section~\ref{sec:methods} describes the CROC simulations, how we measure the halo concentration parameter, and our halo selection strategy. In Section~\ref{sec:results}, we show the concentration-mass relation of the CROC halos, and how baryonic physics implemented in CROC modifies our halo concentrations. In addition, we study the dependence of halo concentration on  the cosmic reionization history.

\section{Methodology}\label{sec:methods}

\subsection{CROC Simulations}

In this study, we use simulations from the Cosmic Reionization On Computers (CROC) project. CROC is a suite of cosmological hydrodynamic simulations of cosmic reionization. Details of the simulations are described in the CROC methods paper \citep{gnedin14}. The sets of CROC simulations we refer to in this paper have simulation volumes of $40 \,h^{-1} {\rm Mpc}$ comoving (cMpc) and spatial resolution of $100 \dim{pc}$ in proper units {\citep[defined as the scale over which the matter density is smoothed,][]{Gnedin2016}} this setup allows CROC to model both the global reionization process and the internal properties of galaxies. For our project, we use three independent random realizations of the $40 \,h^{-1} {\rm cMpc}$ simulation box.

In addition, each of the CROC simulations has a counterpart dark-matter-only simulation that was run using the same initial condition and mass resolution ($7\times10^6\Msun$). These dark-matter-only simulations allow us to compare dark matter halo properties in different environments, and therefore, explore how baryonic physics affects the structure of dark matter halos in CROC. Thus, for our project, we also use data from three dark-matter-only simulations for comparison with the full-physics runs. In the subsequent sections, we refer to the full-physics, hydrodynamic simulations as the ``CROC simulations'', and their dark-matter-only counterparts as the ``DMO simulations".

In both the CROC simulations and the DMO simulations, we use the Rockstar \citep{rockstar} halo finder to identify dark matter halos.

\subsection{Measuring Halo Concentrations}

In Rockstar, $r_{\mathrm{s}}$ is measured assuming that the dark matter halo density profiles follow NFW. However, the NFW profile is originally obtained from halo density profiles in dark-matter-only simulations. We want to properly define the concentration parameter for halos in our full-physics simulations as well, so the first step we take is to study the density profiles of our CROC halos and compare them with the NFW profile. To better visualize the difference when it is not large, we plot the halo circular velocity profiles instead. 

\begin{figure*}[htb!]
    \centering
        \includegraphics[width=\textwidth]{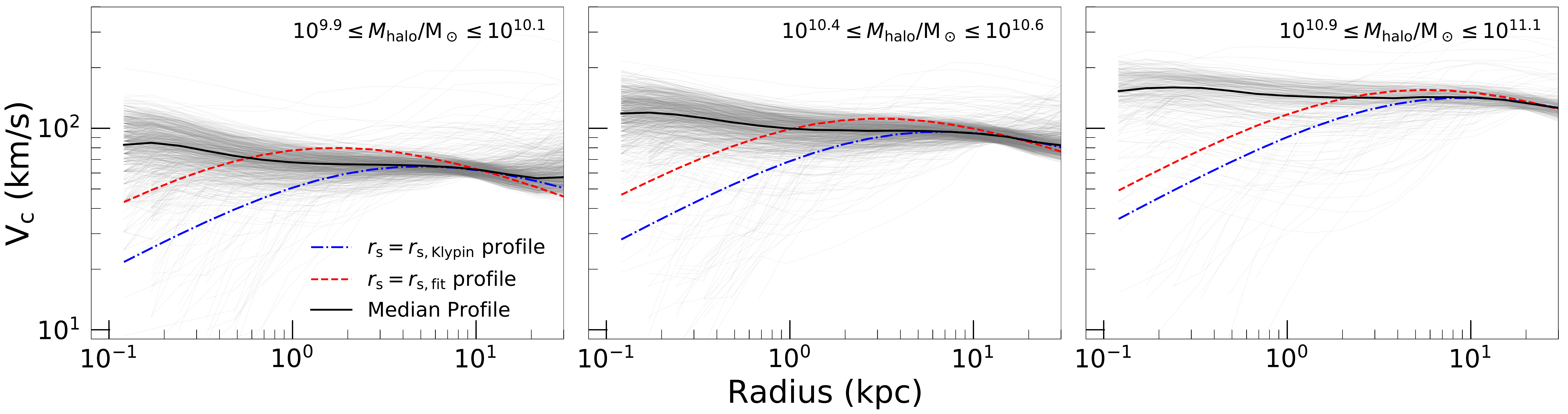} 
\caption{Circular velocity profiles of CROC halos at $z\approx5$ with virial masses in the range $10^{9.9}\Msun < M_{\rm halo} < 10^{10.1}\Msun$, $10^{10.4}\Msun < M_{\rm halo} < 10^{10.6}\Msun$, and $10^{10.9}\Msun < M_{\rm halo} < 10^{11.1}\Msun$, respectively, from left to right.  Thin grey lines represent individual halo profiles, and their median is shown in a thick black line. The individual profiles have been downsampled by a factor of 8 for the $10^{10}\Msun$ halos, and by a factor of 2 for the $10^{10.5}\Msun$ halos.  Red dashed and blue dot-dash lines show analytical NFW profiles calculated by using median values of $r_{\mathrm{s, fit}}$ and $r_{\mathrm{s, Klypin}}$ as found by Rockstar, illustrating that these halos are not well-fit by NFW profiles. } \label{fig:vc_profile}
\end{figure*}

In Figure~\ref{fig:vc_profile}, we show the individual circular velocity profiles $V_c(r)$ and the median circular velocity profile of CROC halos at $z\approx5$ within 3 mass ranges - $10^{9.9}\Msun \leq M_{vir} \leq 10^{10.1}\Msun$, $10^{10.4}\Msun \leq M_{vir} \leq 10^{10.6}\Msun$, and $10^{10.9}\Msun \leq M_{vir} \leq 10^{11.1}\Msun$. The individual circular velocity profiles are shown in solid grey lines, and their median is shown in a solid black line. We calculate the circular velocity profiles simply using
\begin{equation}
V_c(r) = \sqrt{\frac{GM(<r)}{r}}.
\end{equation}
We determine the mass interior to some radius $R$, $M(<R)$, by summing all dark matter particles within $R$ from the center of the dark matter halo, and then we divide the total dark matter mass by the universal dark matter fraction $(\Omega_m - \Omega_b)/\Omega_m$ (0.84 for CROC) to find the total halo mass interior to $R$. This rescaling enables direct comparison with the Rockstar outputs, which we discuss below. For most of the simulated galaxies circular velocity profiles are flat. 

In Figure~\ref{fig:vc_profile}, we also overplot the analytical NFW circular velocity profiles for two different values of the scale radius $r_{\mathrm{s}}$ that the Rockstar halo finder outputs for each halo. The first value is obtained by directly fitting an NFW profile to the data to determine the maximum-likelihood fit of $r_{\mathrm{s}}$, which we are going to call $r_{\mathrm{s, fit}}$. The second approach is to find the radius $r_{\rm max}$ where the circular velocity reaches the maximum $v_{\rm max}$ and then compute $r_{\mathrm{s}}$ from $r_{\rm max}$ under the assumption of an NFW profile \citep{klypin11}. The second value is referred to by Rockstar as $r_{\mathrm{s, Klypin}}$. We calculate the NFW circular velocity profiles using the median values of both definitions of the scale radius as returned by Rockstar. We note that Rockstar excludes ``unbound" particles (defined in \citet{rockstar} as particles whose kinetic energy is greater than the absolute value of potential energy) before calculating $r_{\rm max}$. These unbound particles are usually in the outskirts of dark matter halos, so we expect our median CROC circular velocity profile to slightly differ from the calculated profiles using Rockstar outputs at larger radii. By comparing these two lines with the median halo profile (the solid black line), we can clearly see that the CROC median profiles in all 3 mass ranges deviate significantly from NFW in the inner halo regions. In addition, we have looked into the halo circular velocity profiles at higher redshifts. We conclude the CROC halos are not well-described by the NFW profile at $z \geq 5$ (the CROC simulation is only run until $z\approx5$). We note that the NFW profile fits the density profile of our DMO halos rather well, so departure of the CROC density profile from NFW results from the baryonic processes implemented in the simulations.

Since the CROC halos are not well-fitted by the NFW profile, we opt for a profile-independent measure of halo concentration. We measure the halo concentration using $V_{max}/V_{vir}$, the ratio of the maximum circular velocity to the virial velocity. The maximum circular velocity is simply the maximum value of the quantity $\sqrt{GM(<r)/r}$, and the virial velocity is $\sqrt{GM_{vir}/r_{vir}}$. A larger $V_{max}/V_{vir}$ implies a larger concentration.

Finally, we note that since the goal of this paper is to study halo concentrations, for robustness of our results, we look at only halos with over 1000 dark matter particles within their virial radii because more massive halos are better resolved. Our choice of particle number (or halo mass) cut is based on the convergence study of \citet{mansfield21}. CROC simulations are most similar to the Very Small MultiDark Planck (VSMDPL) simulation in terms of the number of particles per box width and the force softening scale. Our cutoff corresponds to a 4\% error on $V_{max}$ in VSMDPL, which is about half the width of the scatter in the CROC $V_{max}/V_{vir}$.

\subsection{Halo Matching}\label{sec:halo_matching}

To study how reionization physics affects the concentration of dark matter halos, we need to compare the halos in a CROC simulation with those in the counterpart DMO simulation. Therefore, we need to first ``match" halos in these two simulations. Physically, the matched halos should be comprised largely of matter that originate from the same locations in the Lagrangian space in two simulations. Lagrangian space positions in the simulations are identified by unique particle IDs that are assigned in the initial conditions: particles that have the same particle IDs originate from the same Lagrangian coordinates. However, the dynamics in the CROC and DMO runs is somewhat different, since the former includes hydrodynamic forces and the latter does not. Hence, the Lagrangian regions of a given halo in two simulations do not match exactly and differ at the borders. In addition, since CROC was only run until $z\approx5$, the simulated halos are still going through rapid mergers at that time and hence their Lagrangian regions evolve rapidly too. Since in cosmological halos most of the mass is located at large radii (i.e.\ around the borders of their Lagrangian regions), the particles comprising the same physical halo may differ in the two simulations, and that difference is expected to be dominated by particles at large distances (i.e.\ the least bound particles). Thus, in order to better match the halos in the two simulations, we use only particles with the highest binding energy; specifically, we use the 20\% most bound particles to identify the physical counterparts of the CROC and DMO halos.

We bijectively match the halos in CROC and DMO. The procedure is as follows. For a CROC halo, we find its 20\% most bound particles. We locate those particles in the DMO simulation using their particle IDs. Then, we identify the halo that contains the most of these particles in the DMO simulation. We repeat the procedure and find the halo in CROC that contains the highest number of the most bound particles of the identified DMO halo. If the CROC halo we identify is the same CROC halo that we start with, then we establish a bijective match between the CROC halo and DMO halo. Using this method, we are able to match 97\% of our CROC halos with over 1000 DM particles.

\begin{figure}[htb!]
    \centering
        \includegraphics[width=0.99\columnwidth]{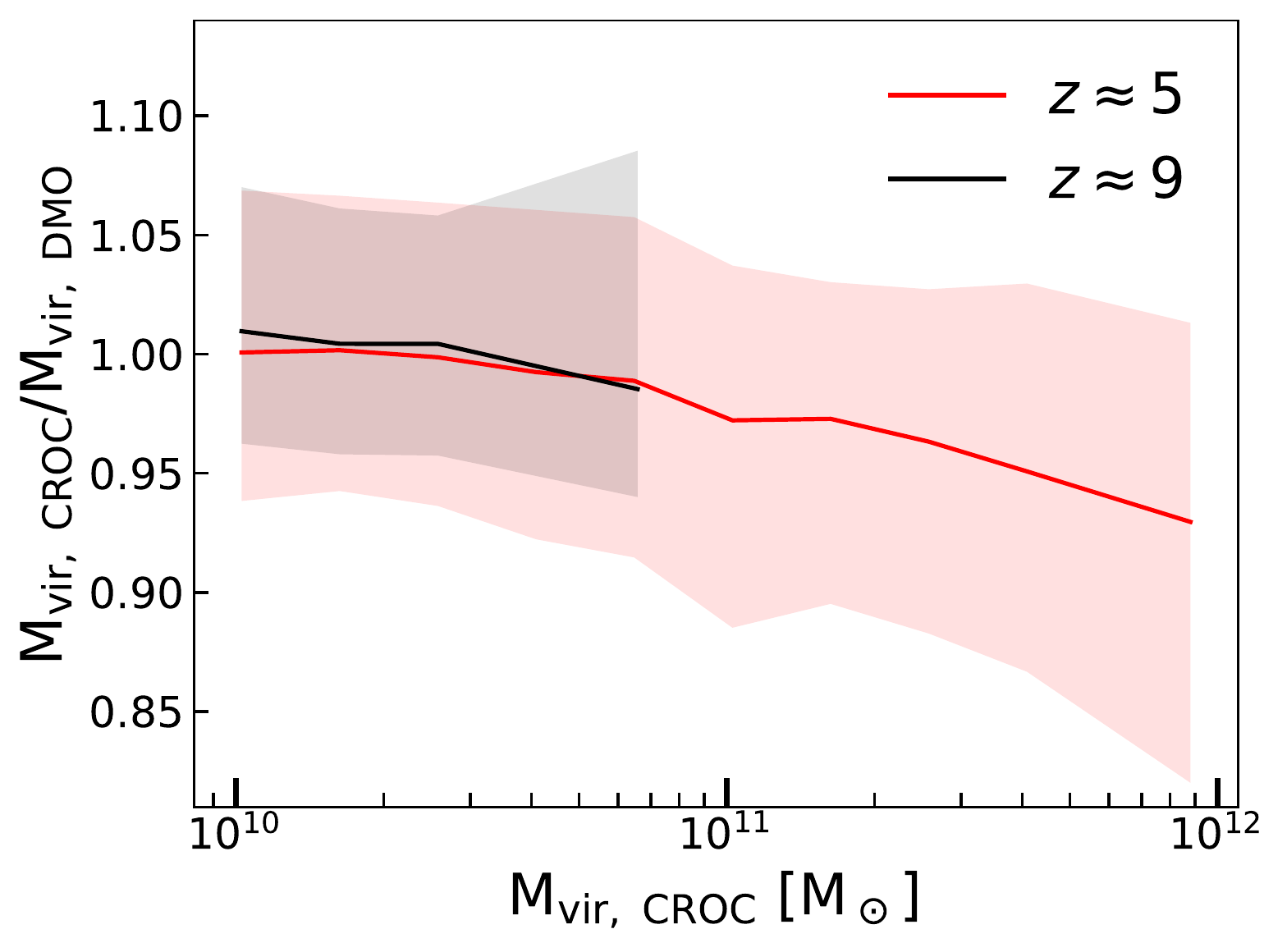} 
\caption{Mass ratio of the matched halos in the CROC and DMO simulations as a function of the CROC halo mass. The black line shows the mass ratio of matched halos at $z\approx9$, and the red line shows the mass ratio at $z\approx5$. CROC halos with masses above about $10^{11}\Msun$ are less massive than their matched DMO halos, but the lower-mass CROC halos have similar masses to those in the DMO runs. Note that halos shown here are too massive to be affected by photo-evaporation.}\label{fig:mass_match}
\end{figure}

Figure~\ref{fig:mass_match} shows the ratio of virial mass ($M_{vir}$) for matched halos in the CROC and DMO simulations as a function of CROC $M_{vir}$. At both $z\approx5$ and $z\approx9$ low-mass CROC halos are similar to the DMO halos, while halos above $10^{11}\Msun$ are slightly less massive than their DMO counterparts. The well-resolved halos we consider here are above the mass scale below which the gas is affected by the photo-evaporation effect \citep[e.g.][]{Okamoto2008}; for lower-mass halos ($M_{vir} \lesssim 10^{9}\Msun$) the DMO masses should exceed the CROC masses again, because these halos are significantly deficient in or totally devoid of gas due to photo-evaporation.

\section{Results}\label{sec:results}
\subsection{The Halo Concentration-Mass Relation}\label{sec:concentration-mass}

\begin{figure*}[htb!]
    \centering
    \begin{minipage}{0.49\textwidth}
        \centering
        \includegraphics[width=0.99\textwidth]{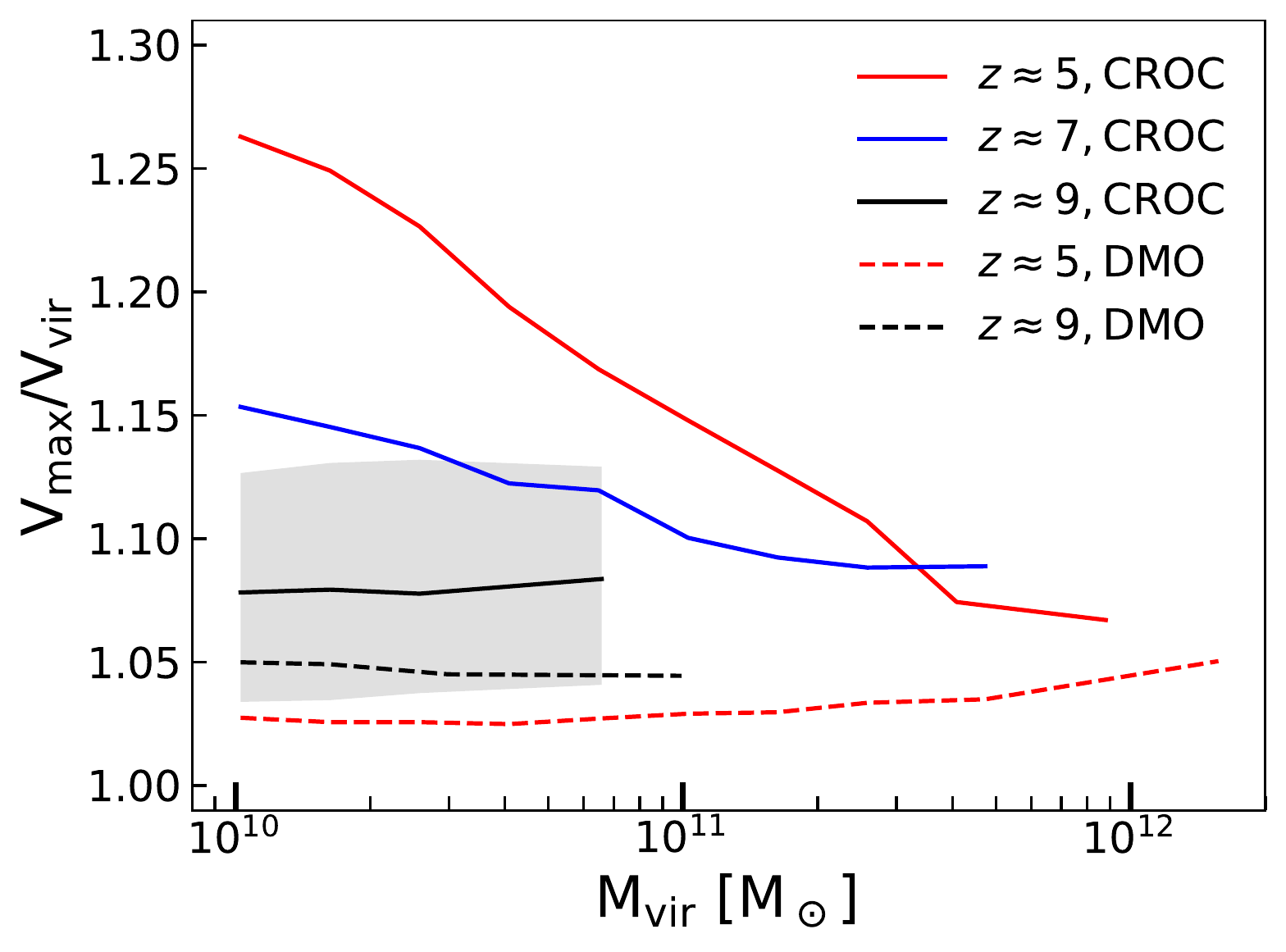} 
    \end{minipage}\hfill
    \begin{minipage}{0.49\textwidth}
        \centering
        \includegraphics[width=0.99\textwidth]{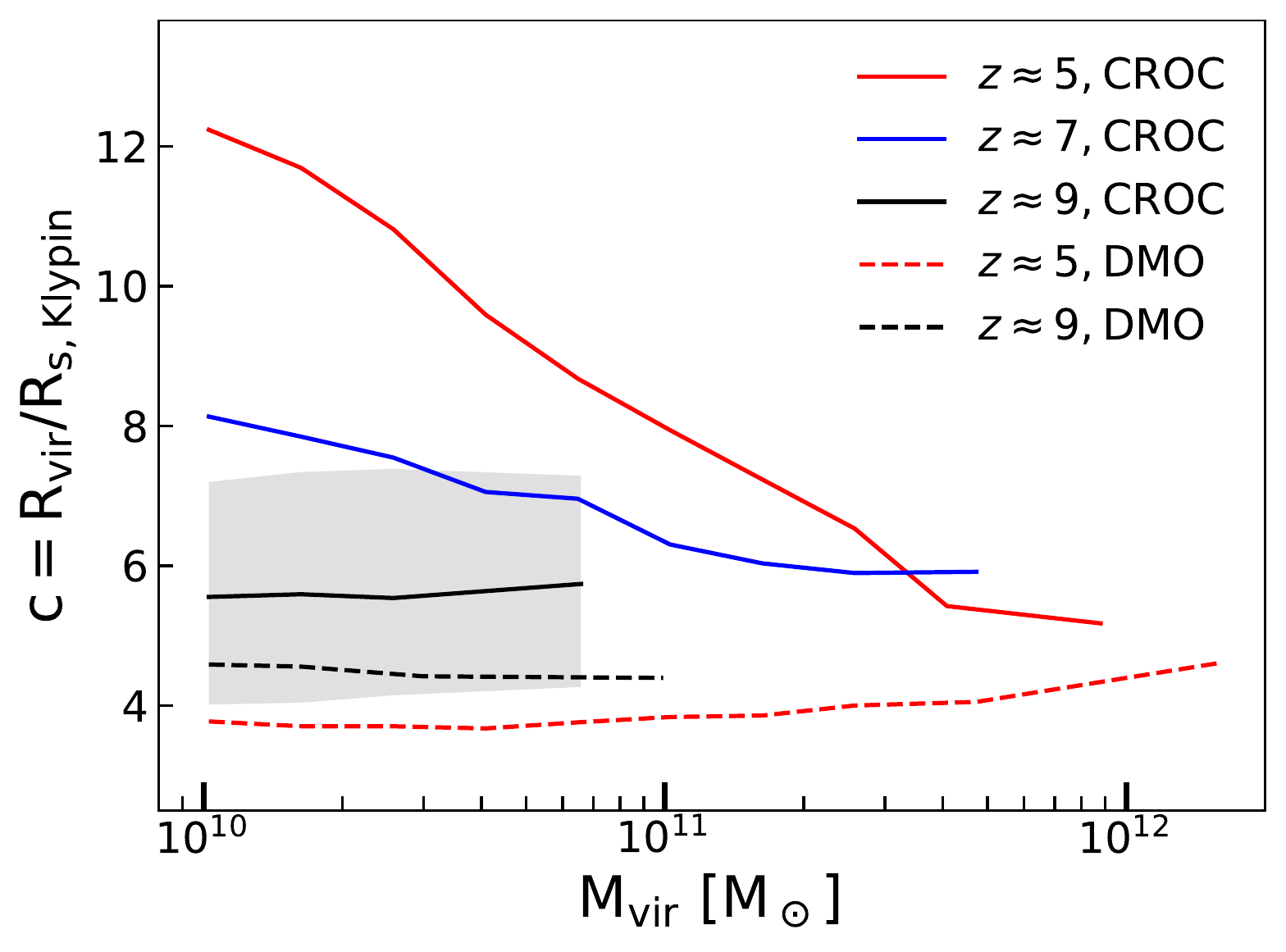} 
\end{minipage}
\caption{Left: $V_{max}/V_{vir}-$mass relation of all halos in CROC and DMO with over 1000 dark mater particles at $z\approx5$, $z\approx7$ (CROC only) and $z\approx9$. The solid lines represent the $V_{max}/V_{vir}-$mass relation of the CROC halos, while the dashed lines show that of the DMO halos. Different line colors correspond to different redshifts. Shaded region shows the 25th to 75th percentile of $V_{max}/V_{vir}$ per mass bin for the $z\approx9$ CROC halos.  All lines have comparable scatter. Right: Concentration-mass relation of the same halos, where the concentration parameter $c$ is defined as the ratio of the halo virial radius to the Klypin scale radius. Because $r_{\mathrm{s, Klypin}}$ is defined via $V_{max}$, the halo concentration defined with $r_{\mathrm{s, Klypin}}$ contains largely the same information as $V_{max}/V_{vir}$.}\label{fig:c-m}
\end{figure*}

Besides causing a small change in the virial mass shown in Figure~\ref{fig:mass_match}, baryonic physics primarily affects the concentration of dark matter halos. Figure~\ref{fig:c-m} summarizes the mass dependence of concentrations in CROC and DMO simulations (with concentrations measured both as $V_{max}/V_{vir}$ and as $R_{vir}/r_{\mathrm{s, Klypin}}$). Each line corresponds to the median
$V_{max}/V_{vir}$ concentration in each mass bin. Lines of the same color show the relation at the same redshift, with solid lines showing the CROC results and dashed lines showing the DMO results. At $z\approx5$ and $z\approx7$ the CROC $V_{max}/V_{vir}$ concentration decreases as mass increases in the mass range of $10^{10}\Msun$ to $10^{12}\Msun$, whereas at $z\approx9$ the concentration is essentially independent of mass in the mass range of $10^{10}\Msun$ to $10^{11}\Msun/h$. For DMO, we find overall lower $V_{max}/V_{vir}$ concentration values as a function of halo mass at both redshifts. There is no measurable trend with mass at $z\approx9$ and only a weak increasing trend of $V_{max}/V_{vir}$ with mass at $z\approx5$. 

In the right panel of Figure~\ref{fig:c-m} we show the CROC and DMO median concentration as a function of halo mass for the same epochs and simulations. The $c-M$ relation has the same trends and redshift dependence as the $V_{max}/V_{vir}-$mass relation. Recall that we define concentration as $R_{vir}/r_{\mathrm{s, Klypin}}$. Since $r_{\mathrm{s, Klypin}}$ is set by the location of $V_{max}$, we expect that both relations contain essentially the same information, unless $V_{max}/V_{vir}$ and $R_{max}/R_{vir}$ are independent. Hence, the right panel of Figure~\ref{fig:c-m} indicates that $V_{max}/V_{vir}$ and $R_{max}/R_{vir}$ are strongly correlated for our halos. This correlation is expected for DMO runs but it is not necessarily the case for full physics runs. For the CROC halos, however, this is indeed the case.

\begin{figure}[htb!]
    \centering
        \includegraphics[width=0.99\columnwidth]{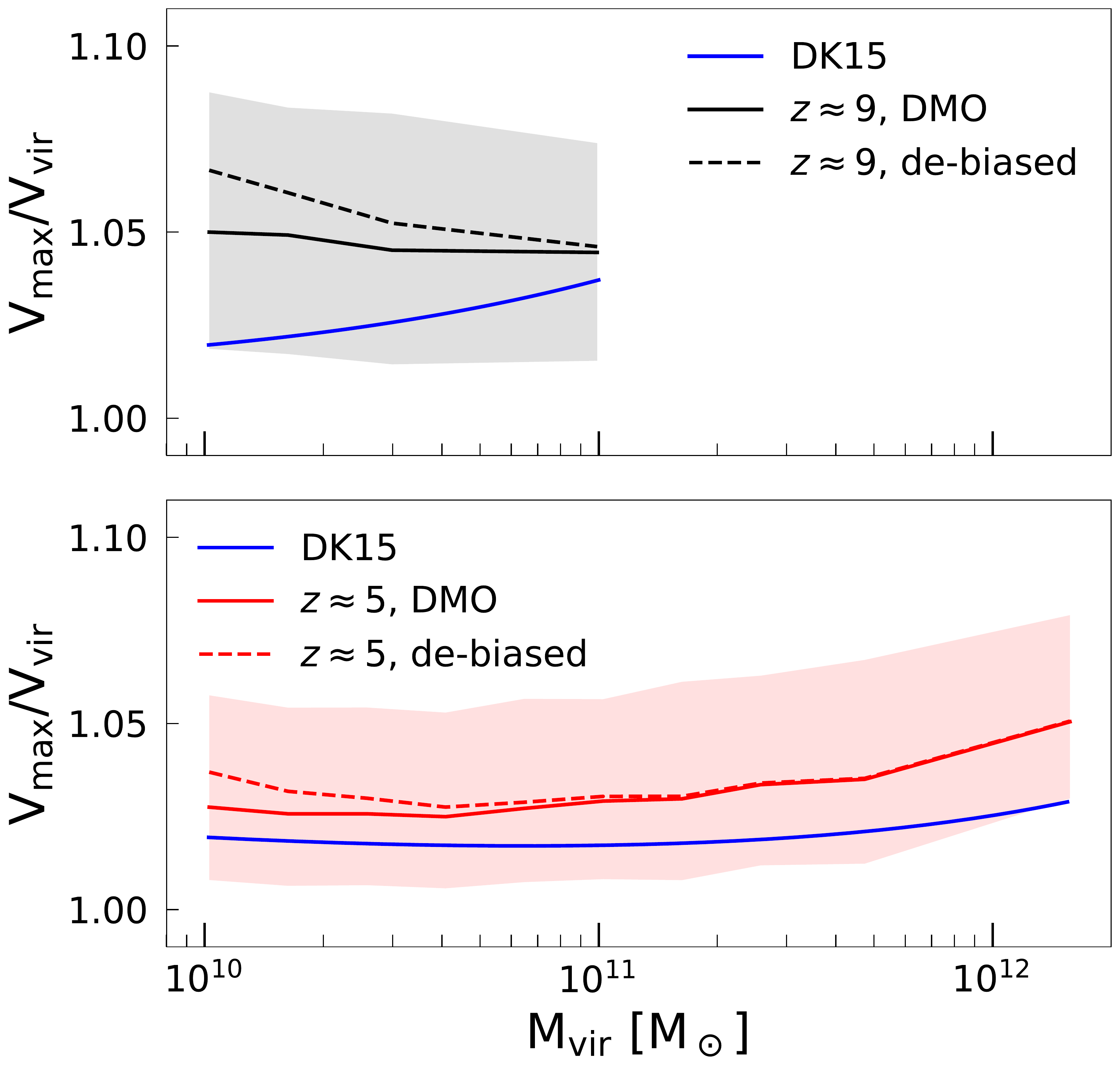} 
\caption{Top: $V_{max}/V_{vir}-$mass relation of DMO halos at $z\approx9$ (solid black), compared with the DK15 model (solid blue) and the ``de-biased" results using the \citet{mansfield21} method (dashed black). Bottom: $V_{max}/V_{vir}-$mass relation of the same DMO halos at $z\approx5$ (solid red), also compared with the DK15 model and the ``de-biased" results (dashed red). Shaded regions indicate the 25th to 75th percentile spread for the DMO relation in both redshifts.  Our DMO results are consistent with the DK15 model within the spread and have sufficient resolution such that the bias correction does not significantly change the relation for most of our mass range.
} \label{fig:dmo_compare}
\end{figure}

One potential concern for measuring concentrations is the spatial resolution of simulations. While CROC simulations have the maximum spatial resolution of 100 pc in proper units, more than sufficient to measure concentrations of all resolved halos, the DMO runs have a lower resolution of only 3.6 comoving kpc, and hence may suffer from numerical effects. To check for that, we follow the method of ``de-biasing" $V_{max}/V_{vir}$ as described in \citet{mansfield21} to correct for the numerical effects in DMO simulations that can bias the profiles of dark matter halos. We plot the de-biased $V_{max}/V_{vir}$ as a function of mass along with the original $V_{max}/V_{vir}$ in Figure~\ref{fig:dmo_compare}. We observe that the de-biased values do not significantly differ from our simulation results at both $z\approx5$ and $z\approx9$. Thus, we conclude that our DMO simulations are suitable for this study. 

Another important check for the accuracy of our presented results is the comparison with well-established results. To that end we also plot in Figure~\ref{fig:dmo_compare} the \citet{DK15} model (hereafter DK15) for the concentration-mass relation of dark matter halos at $z\approx5$ and $z\approx9$ for comparison (note that we convert their modeled concentration values to $V_{max}/V_{vir}$ values assuming NFW). DK15 proposed a method of modeling concentration as a function of peak height and the logarithmic power spectrum slope. We notice that CROC has slightly higher $V_{max}/V_{vir}$ concentration values than the DK15 model at both redshifts. At $z\approx5$ the difference is really negligible. At $z\approx9$ our simulated values for $V_{max}/V_{vir}$ are essentially mass-independent while the DK15 model predicts a mild trend with mass; however, the DK15 prediction remains well within the scatter of the simulation results. We deem this level of agreement sufficient for our purposes.

\subsection{CROC and DMO Comparison}

\begin{figure}[htb!]
    \centering
        \includegraphics[width=0.99\columnwidth]{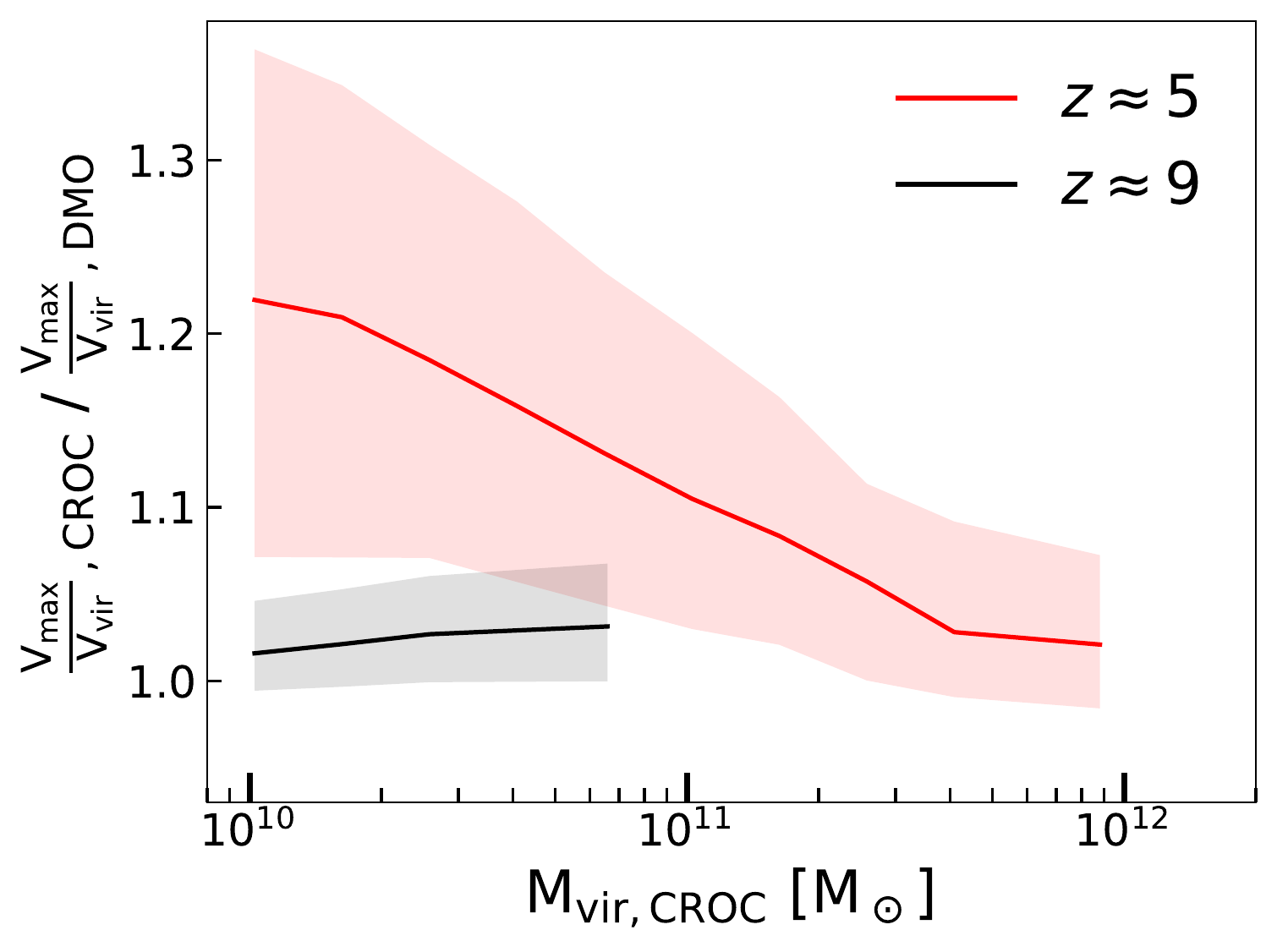} 
\caption{$V_{max}/V_{vir}$ ratio of matched CROC halos to DMO halos at $z\approx5$ (in red) and $z\approx9$ (in black). Shaded regions show the 25th to 75th percentiles of the quantities. At $z\approx5$, the $V_{max}/V_{vir}$ ratio decreases with halo mass, whereas $z\approx9$ shows a weak opposite mass trend. } \label{fig:ratio_z}
\end{figure}

With numerical effects under control, we now compare the CROC and DMO $V_{max}/V_{vir}$ concentrations on a halo-by-halo basis, using the matched halos as described in Section~\ref{sec:halo_matching}. This comparison isolates the effects of baryons on individual halos from the general scatter in the distributions. Figure~\ref{fig:ratio_z} shows the ratio of CROC $V_{max}/V_{vir}$ to DMO $V_{max}/V_{vir}$ between individually matched halos. At both redshifts the $V_{max}/V_{vir}$ concentration is increased in the presence of baryons, and that is fully expected as gas cooling results in baryons concentrating at the halo center. At $z\approx9$, the very onset of structure formation, this effect is minor, as most of the gas in a typical halo we resolve ($M_{vir}\sim10^{10.5}\Msun$) would not have time to cool significantly. At $z\approx5$ the increase in the concentration ratio is very significant at low halo mass and this ratio decreases with halo mass. Since cooling times are weakly dependent on halo mass (and, if anything, may be shorter in more massive, hotter halos in the presence of strong radiation fields), the decrease in the ratio of halo concentrations can only be due to the effect of stellar feedback, which expels baryons from halo centers and reduces halo concentrations. It is important to note here that the stellar feedback is overestimated in CROC simulations in halos with $M_{vir}\gtrsim10^{11}\Msun$ \citep{zhu20}, hence the rate of decrease in the CROC-to-DMO ratio of concentrations may be overestimated for the most massive halos.

\begin{figure}[htb!]
    \centering
        \includegraphics[width=0.99\columnwidth]{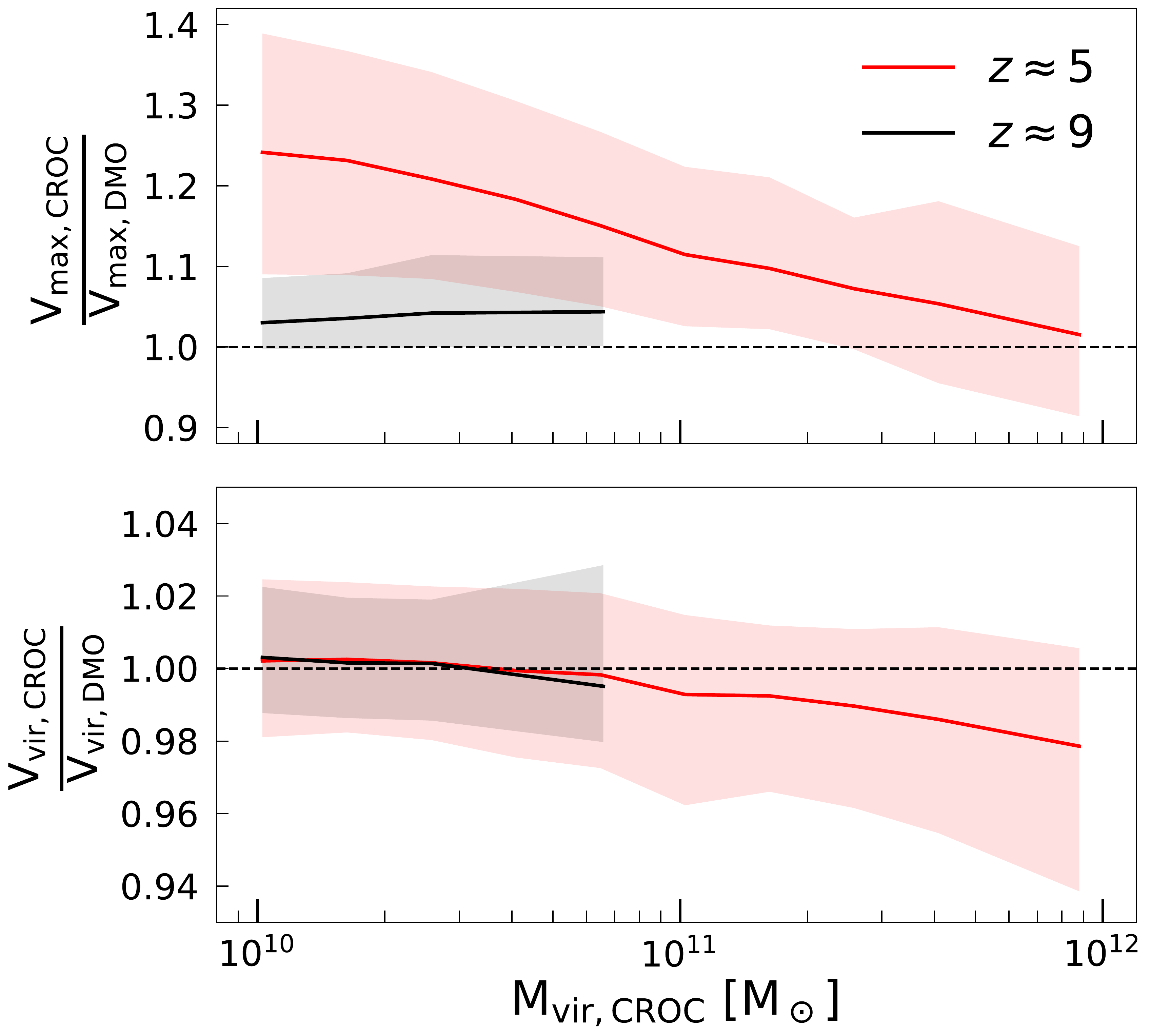} 
\caption{Top: $V_{max}$ ratio between the matched halos in CROC and DMO at $z\approx5$ and $z\approx9$. Bottom: $V_{vir}$ ratio between matched halos in CROC and DMO at $z\approx5$ and $z\approx9$. Consistent with the color choices throughout the paper, red is for $z\approx5$ and black is for $z\approx9$. Shaded regions denote the 25th to 75th percentile. If all the baryons were simply centrally concentrated, the ratio in the top panel would be $\approx 2.5$.  The reduced ratio indicates that feedback processes control the halo mass distribution for all halos.} 
\label{fig:ratio_sep_z}
\end{figure}

The ratio of concentration depends on the combined effect of gas cooling and stellar feedback. In order to separate the two effects, we show in Figure~\ref{fig:ratio_sep_z} contributions to the $V_{max}/V_{vir}$ ratio from $V_{max}$ and $V_{vir}$ separately. In the top panel of Figure~\ref{fig:ratio_sep_z}, we plot the ratio of $V_{max}$ between the matched CROC halos and DMO halos; in the bottom panel, we show the ratio of $V_{vir}$ between the matched halos. If gas cooling dominated the baryon dynamics for some halos (i.e.\ if all the baryons were simply pulled towards the center), the increase in halo $V_{max}$ in the top panel of Figure~\ref{fig:ratio_sep_z} should be a factor of about 2.5 \citep{Gnedin2002}. This is much larger than the increase we actually measure in the simulations, so the role of feedback is dominant in shaping the mass profiles of our simulated galaxies for all halo masses.

\subsection{Environmental Dependence of Halo Concentration}

While stellar feedback is ``local" on cosmic scales (i.e.\ affecting primarily the host halos), the radiative feedback is global - radiation emitted by a galaxy may affect other galaxies many megaparsecs away. Since CROC simulates cosmic reionization, it is useful to explore the effect of the global radiative feedback (i.e.\ the reionization history) on halo concentrations.

The CROC simulations track the neutral hydrogen fractions ($x_{\rm HI}$) on the grid. \citet{gnedin14} and \citet{gnedinandkaurov_14} have explored the time evolution of the neutral fraction throughout the epoch of reionization. To distinguish between environments that experience reionization at different times, we divide each of our simulation boxes into smaller, equal-size regions with their average $x_{\rm HI}$ values at a fixed redshift as indicators of the reionization times. In the simulation suites we use for this study, the average volume-weighted $x_{\rm HI}$ over the entire simulation box is 0.5 at $z\approx8$. So we opt to use the local average $x_{\rm HI}$ values at $z\approx8$ as a proxy for timing of reionization in different environments. Since the typical scale of ionized bubbles is about $10 \,h^{-1} {\rm cMpc}$, we divide our $40 \,h^{-1} {\rm cMpc}$ simulation box into 64 equal-volume cubes, so each has a length of $10 \,h^{-1} {\rm cMpc}$. With periodic boundary conditions such a division is not unique, so we move the origin to maximize the ratio of $\max(\bar{x}_{\rm HI})/\min(\bar{x}_{\rm HI})$ among the 64 sub-boxes, where $\bar{x}_{\rm HI}$ is the average value of the hydrogen neutral fraction in each sub-box. We follow this procedure to divide all three simulation boxes as each of them is an independent random realization of our universe. We list the average $\bar{x}_{\rm HI}$  values in the most and least ionized sub-boxes in Table~\ref{table:average_xHI}.

\begin{figure*} [htb!]
    \centering
        \includegraphics[width=\textwidth]{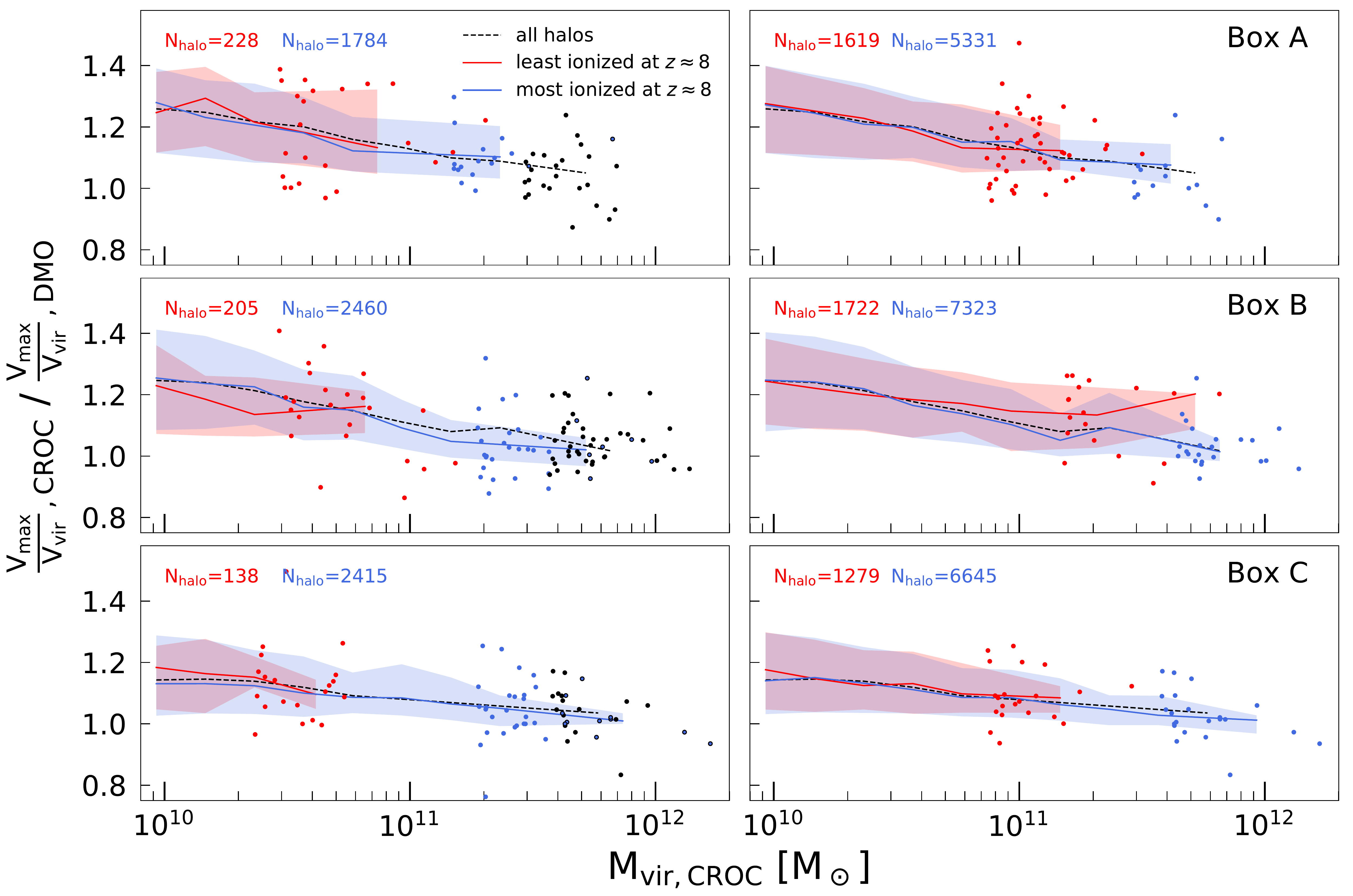}
\caption{$z\approx5$ $V_{max}/V_{vir}$ ratio between matched halos in CROC and DMO in environments separated by reionization histories characterized by neutral fraction at $z\approx8$. Out of 64 sub-boxes in each simulation box, the left panels show the ratio in the 4 most (red) and least (blue) ionized $10 \,h^{-1} {\rm cMpc}$ sub-boxes. The right panels show the ratio in the 16 most and least ionized $10 \,h^{-1} {\rm cMpc}$ sub-boxes. From top to bottom, we show the results from 3 independent random realizations of our $40 \,h^{-1} {\rm cMpc}$ simulation box. In each subplot, we also show the $V_{max}/V_{vir}$ ratio for all of the matched halos in the corresponding simulation box (in black). In mass bins where there are few halos we show individual halos as dots of the corresponding color. We find that the reionization histories do not significantly affect the $V_{max}/V_{vir}$ concentration of halos.} \label{fig:compare_xHI}
\end{figure*}

We then study how reionization physics affects the $V_{max}/V_{vir}$ measure of the concentration of galaxies living in environments with different reionization histories. To obtain better halo statistics, in each box we group the 4 most ionized regions at $z\approx8$ together, and the 4 least ionized regions at $z\approx8$ together. We also show the results after grouping the 16 most ionized cubes and 16 least ionized cubes for comparison. The average neutral fractions $\bar{x}_{\rm HI}$ in these groups are listed in Table~\ref{table:average_xHI} as well. In Figure~\ref{fig:compare_xHI} we show the halo-to-halo comparison of the $V_{max}/V_{vir}$ concentration in CROC and DMO at $z\approx5$, when much of the universe has been reionized. The three panels on the left show the ratio of $V_{max}/V_{vir}$ between matched CROC and DMO halos living in the 4 most and least ionized $10 \,h^{-1} {\rm cMpc}$ sub-boxes in boxes A, B and C, respectively, from top to bottom. The three panels on the right compare the $V_{max}/V_{vir}$ ratio of CROC and DMO halos living in the 16 most and least ionized sub-boxes. In all 6 panels, we use red color to show halos living in the least ionized environments, and blue color for halos in the most ionized environments. Additionally, we plot the ratio for all halos in each box in dashed black lines for comparison. In mass bins with few halos, we show individual halos as dots. We find that the reionization histories do not significantly affect the $V_{max}/V_{vir}$ concentration of halos at $z\approx5$. Note that there are fewer galaxies living in the least ionized environments (with large $\bar{x}_{\rm HI}$, shown in red in Figure~\ref{fig:compare_xHI}). This corresponds to the ``inside-out" phase of cosmic reionization where dense regions get reionized first.

\begin{table}[htb]
\centering
\begin{tabular}{c c c c}  
\hline
$\bar{x}_{\rm HI}$ & Box A & Box B & Box C \\
\hline
$\max(\bar{x}_{\rm HI}|\mbox{1 box})$ & 0.89& 0.83& 0.98\\
$\min(\bar{x}_{\rm HI}|\mbox{1 box})$ &  0.13 & 0.02 & 0.23\\
$\max(\bar{x}_{\rm HI}|\mbox{4 boxes})$ & 0.84 & 0.76 & 0.98\\
$\min(\bar{x}_{\rm HI}|\mbox{4 boxes})$ & 0.21 & 0.05 & 0.39\\
$\max(\bar{x}_{\rm HI}|\mbox{16 boxes})$ & 0.76 & 0.60 & 0.94\\
$\min(\bar{x}_{\rm HI}|\mbox{16 boxes})$ & 0.32 & 0.13 & 0.57\\
\hline
\end{tabular}
\caption{Average neutral hydrogen fraction, $\bar{x}_{\rm HI}$, in our regions of interest. From top to bottom: $\bar{x}_{\rm HI}$ in the least ionized $10 \,h^{-1}{\rm cMpc}$ sub-box, $\bar{x}_{\rm HI}$ in the most ionized $10 \,h^{-1}{\rm cMpc}$ sub-box, $\bar{x}_{\rm HI}$ in the 4 least ionized $10 \,h^{-1}{\rm cMpc}$ sub-boxes, $\bar{x}_{\rm HI}$ in the 4 most ionized $10 \,h^{-1}{\rm cMpc}$ sub-boxes, $\bar{x}_{\rm HI}$ in the 16 least ionized $10 \,h^{-1}{\rm cMpc}$ sub-boxes, $\bar{x}_{\rm HI}$ in the 16 most ionized $10 \,h^{-1}{\rm cMpc}$ sub-boxes.} \label{table:average_xHI}
\end{table}

\section{Summary and Discussion}

We have used simulations from the Cosmic Reionization On Computers (CROC) project and their counterpart dark-matter-only (DMO) simulations to study the effect of baryonic physics that CROC models on the concentration of dark matter halos at $5 \leq z \leq 9$.

Our main results are:
\begin{enumerate}
\item The CROC halo density profiles at $z \geq 5$ are not well-fitted by the NFW profile. We therefore use the ratio, $V_{max}/V_{vir}$, for a more robust characterization of the CROC halo concentration.

\item At $z\approx5$ and $z\approx7$, the CROC halo concentration decreases with mass, whereas at our highest redshift snapshot, $z\approx9$, concentration is mass independent. (See Figure~\ref{fig:c-m}.)

\item At both $z\approx5$ and $z\approx7$, baryons increase the $V_{max}/V_{vir}$ concentration of CROC halos. At $z\approx5$, the lower mass halos in our simulations are  more affected by the baryonic physics. At $z\approx9$, however, the more massive halos are more affected, although the overall effect is small. (See Figure~\ref{fig:ratio_z}.)

\item The effect of the baryonic physics on halo concentrations that we measure in the simulations is much less than what would be expected if all the baryons were simply pulled towards the center \citep{Gnedin2002}. Because cooling is very rapid at these redshifts, it must be the stellar feedback that controls the spatial extent of baryons inside galactic halos and, hence, the halo concentrations that we measure. (See Figure~\ref{fig:ratio_sep_z}).

\item Reionization history, characterized by the neutral fraction at $z=8$, does not have a significant impact on the CROC halo concentration at $z\approx 5$. (See Figure~\ref{fig:compare_xHI}.)

\end{enumerate}

We note that one can characterize the reionization history of halos in multiple ways \citep{zhu19}. This, in turn, may illustrate varying relationships with the impact of baryonic physics on concentrations and other properties at high redshifts.  However, we leave studies in this direction for a follow-up paper.

\acknowledgments

The authors thank Camille Avestruz and Phil Mansfield for helpful discussions which improved this work. Fermilab is operated by Fermi Research Alliance, LLC, under Contract No. DE-AC02-07CH11359 with the United States Department of Energy. This work used resources of the Argonne Leadership Computing Facility, which is a DOE Office of Science User Facility supported under Contract DE-AC02-06CH11357. An award of computer time was provided by the Innovative and Novel Computational Impact on Theory and Experiment (INCITE) program. This research is also part of the Blue Waters sustained-petascale computing project, which is supported by the National Science Foundation (awards OCI-0725070 and ACI-1238993) and the state of Illinois. Blue Waters is a joint effort of the University of Illinois at Urbana-Champaign and its National Center for Supercomputing Applications.

\bibliographystyle{apj}
\bibliography{main}

\end{CJK*}
\end{document}